\documentclass[twocolumn,showpacs,preprintnumbers,amsmath,amssymb,prl]{revtex4}

\usepackage{graphicx}
\usepackage{dcolumn}

\usepackage[centerlast,footnotesize]{caption}

\usepackage[latin1]{inputenc}
\usepackage[T1]{fontenc}
\usepackage{ae,aecompl}

\begin{document}
\captionsetup{labelsep= none,
  justification=centerlast,width=.9\columnwidth,aboveskip=5pt}

\newcommand{\rar}{$\rightarrow$}
\newcommand{\lrar}{$\leftrightarrow$}

\newcommand{\beq}{\begin{equation}}
\newcommand{\eeq}{\end{equation}}
\newcommand{\bea}{\begin{eqnarray}}
\newcommand{\eea}{\end{eqnarray}}
\newcommand{\Req}[1]{Eq. (\ref{E#1})}
\newcommand{\req}[1]{(\ref{E#1})}
\newcommand{\degree}{$^{\rm\circ} $}
\newcommand{\pcite}{\protect\cite}
\newcommand{\pref}{\protect\ref}
\newcommand{\Rfg}[1]{Fig. \ref{F#1}}
\newcommand{\rfg}[1]{\ref{F#1}}
\newcommand{\Rtb}[1]{Table \ref{T#1}}
\newcommand{\rtb}[1]{\ref{T#1}}

%

\title{Anharmonic Torsional Stiffness of DNA Revealed under Small
External Torques}

\author{Alexey K. Mazur}
\email{alexey@ibpc.fr}
\affiliation{CNRS UPR9080, Institut de Biologie Physico-Chimique,
13, rue Pierre et Marie Curie, Paris,75005, France.}
 


\begin{abstract}
DNA supercoiling plays an important role in a variety of cellular
processes. The torsional stress related with supercoiling may be also
involved in gene regulation through the local structure and dynamics
of the double helix. To check this possibility steady torsional stress
was applied to DNA in the course of all-atom molecular dynamics
simulations. It is found that small static untwisting significantly
reduces the torsional persistence length ($l_t$) of GC-alternating
DNA.  For the AT-alternating sequence a smaller effect of the opposite
sign is observed. As a result, the measured $l_t$ values are similar
under zero stress, but diverge with untwisting. The effect is traced
to sequence-specific asymmetry of local torsional fluctuations, and it
should be small in long random DNA due to compensation. In contrast,
the stiffness of special short sequences can vary significantly, which
gives a simple possibility of gene regulation via probabilities of
strong fluctuations. These results have important implications for
the role of local DNA twisting in complexes with transcription
factors.
\end{abstract}


\maketitle

The double helical DNA in living cells is subjected to a constitutive
unwinding torque created by special enzymes. This forces DNA to fold in
a supercoiled state similarly to a flexible rod with bending and
twisting elasticity. The supercoiling is long known to play an
important role in a variety of cellular processes
\cite{Vologodskii:94b}. Its magnitude changes regularly during the
cell cycle and in response to environmental conditions, which is
accompanied by activation or suppression of certain genes
\cite{Travers:05a}. In E.  coli, relaxation of the superhelical stress
simultaneously alters activity of 306 genes (7\% of the genome), with
106 genes activated and other deactivated \cite{Peter:04}. The genes
concerned are functionally diverse, widely dispersed throughout the
chromosome, and the  effect is dose-dependent. These and many similar
observations suggest that the DNA supercoiling is used as a universal
transcriptional regulator \cite{Travers:05a}, but the corresponding
physical mechanisms are not clear.

Detailed studies indicate that the promoter sensitivity to supercoiling
stems from the recognition of promoter elements by RNA polymerase,
and that it does not require DNA melting or transitions to alternative
forms \cite{Borowiec:87}. The supercoiling torque is distributed
between twisting and writhing so that the untwisting of the double
helix is estimated as 1-2\% \cite{Boles:90}, which is below the
thermal noise and too small for reliable recognition. However, the
action of the torsional stress can be conveyed through a property
rather than the structure  of the double helix. The behavior of the
supercoiled DNA is governed by the interplay between the local bending
and twisting fluctuations. If the bending flexibility or the torsional
stiffness of the double helix vary with forced untwisting, parameters
of thermal fluctuations could be noticeably affected already for short
DNA stretches involved in the recognition. This idea is appealing and
it is supported by some earlier data for long DNA
\cite{Song:90b,Selvin:92,Naimushin:94}. Local torsional fluctuations
are likely to be involved in regulation directly. In bacterial
promoters, the optimal linker between the -10 and -35 elements
involves 16 base pair steps (bps), but in promoters sensitive to
supercoiling it is usually one step shorter or longer
\cite{Borowiec:87,Jordi:95}. One step corresponds to rotation by
34.5$^\circ$, which approximately equals the root-mean square width of
torsional fluctuations for the linker.  Very strong torsional
fluctuations of short DNA stretches are necessary for activation of
some animal promoters \cite{Chow:91}.

Local effects of the torsional stress are difficult to reveal
experimentally, but they can be probed by all-atom MD simulations. New
methods were recently developed to apply steady forces and torques to
short stretches of DNA \cite{Mzjctc:09}. In contrast to twisting by
periodic boundary constraints and potential restraints used earlier
\cite{Kannan:06,Wereszczynski:06,Randall:09} the steady stress
emulates local conditions of a short fragment in a long supercoiled
DNA, which makes possible evaluation of elastic parameters under very
low torsional load corresponding to physiological conditions.  This
method captures linear elastic responses as well as the twist-stretch
coupling effect under small torques corresponding to physiological
degree of supercoiling \cite{Mzjctc:09}. Here we present the results
of the first computational study of the elastic parameters of DNA in
such conditions.

Dynamics of two tetradecamer DNA with AT- and GC-alternating
sequences, respectively, were simulated in explicit aqueous solution
using earlier described protocols \cite{Mzjctc:09}. For each
duplex, nine 164 ns trajectories of all-atom dynamics were computed
with fixed torque values in the range $\pm$ 20 pN$\cdot$nm, which
gives about 3 $\mu$s of simulations in total. Three additional
trajectories were computed for the GC-alternating fragment for
verification. Below we consider only evaluation of the torsional
stiffness. Other methods and protocols are described in
Appendix. In the harmonic approximation the torsional free energy
of a DNA fragment of length $L$ subjected to external torque
$\tau$ is
\beq\label{EU}
U(\Phi)=kT\frac{l_t}{2L}\left(\Phi-\Phi_\tau\right)^2
\eeq
where $\Phi$ is the overall winding angle,  $\Phi_\tau$ is its
equilibrium value, $l_t$ is the torsional persistence length, and
$kT$ is the Boltzmann's factor. The equilibrium winding varies with
the torque as
\beq\label{EPhi}
\Phi_\tau-\Phi_0=\frac{\tau L}{kTl_t}
\eeq
In the course of MD simulations one measures the probability
distribution $P_{\Phi}$ for the winding angle of one helical turn
which, in the limit of infinite sampling, has a canonical form
\beq\label{EPphi}
P_{\Phi}\sim\exp\left[-\frac{l_t}{2L}\left(\Phi-\Phi_\tau\right)^2\right].
\eeq
The equilibrium winding is estimated as the time average
$\langle\Phi\rangle_t$, and the torsional persistence length $l_t$ is
extracted from the time variance $\Delta^2_t\Phi$. The potential of
mean force (PMF) corresponding to any Gaussian distribution is
quadratic, but if the harmonic approximation is truly valid, $l_t$
must be constant with different $\tau$.

\begin{figure}[ht]
\centerline{\includegraphics[width=7cm]{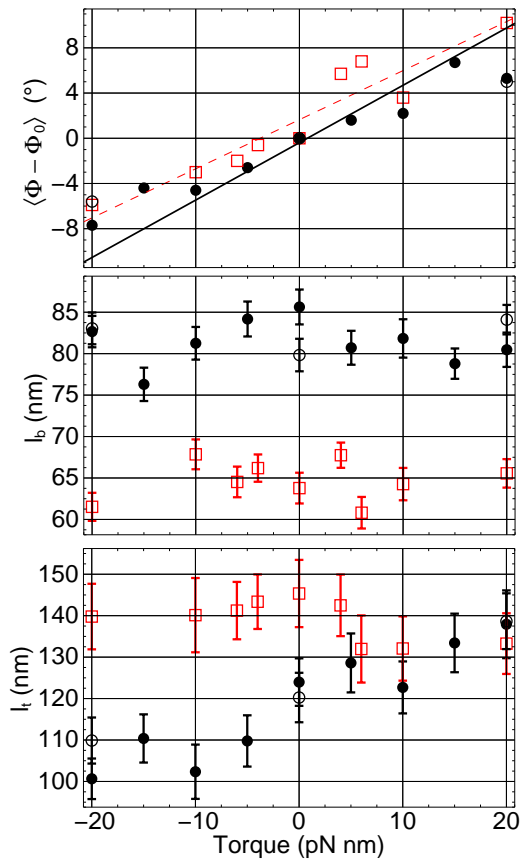}}
\caption{\label{Falvstq} Color online.
Representative torque dependences obtained by all-atom MD simulations.
The results are shown for the overall twisting (top panel), the
bending persistence length ($l_b$, middle panel), and the torsional
persistence length (bottom panel) of the AT-alternating
($\Phi_0\approx363.1^\circ$, red squares) and GC-alternating
($\Phi_0\approx381.8^\circ$, black circles) sequences. The open
circles feature the verification tests. The straight lines on the top
panel correspond to \Req{Phi} with $l_t=124$ nm (solid black line) and
$l_t=145$ nm (dashed red line). The error bars show statistical errors
evaluated by the method of block averages (see Appendix). In the top
panel the symbol size corresponds to maximal errors.
}\end{figure}

The top panel of \Rfg{alvstq} shows variations of $\Phi_\tau$
corresponding to \Req{Phi}. All measurements were taken  for the
central 12 bp stretches, with the two terminal steps ignored, which
gives about one helical turn. The amplitude of the forced winding is
$\pm$ 2 \%, i.e. about 0.7$^\circ$ per base pair. The straight lines
shown have the slopes corresponding to $l_t$ obtained under zero
torque. In the range of torques $\pm$ 10 pN$\cdot$nm the points are
compatible with a linear elastic response (harmonic elasticity).
Beyond this range the profile remains roughly linear for the
AT-alternating sequence, but for the GC-alternating duplex evident
deviations from harmonicity are found.  These deviations are
reproducible and quite strong. If the $l_t$ value were evaluated by
\Req{Phi} using $\Phi_\tau$ for $\tau=\pm20$ pN$\cdot$nm it would be
about 200 nm.

The measured torsion persistence length changes with the applied
torque as shown in the bottom panel.  The GC-alternating sequence
exhibits strong anharmonicity, with the twist increase of 1.4$^\circ$
per bps accompanied by 30\% growth in $l_t$.  For the AT-alternating
sequence, the $l_t$ profile is nearly flat with a small decreasing
trend. This trend becomes more visible with stronger twisting (article
in preparation).  The bending stiffness varies somewhat beyond the
estimated statistical errors, but without regular trends.

\begin{figure}[ht]
\centerline{\includegraphics[width=7cm]{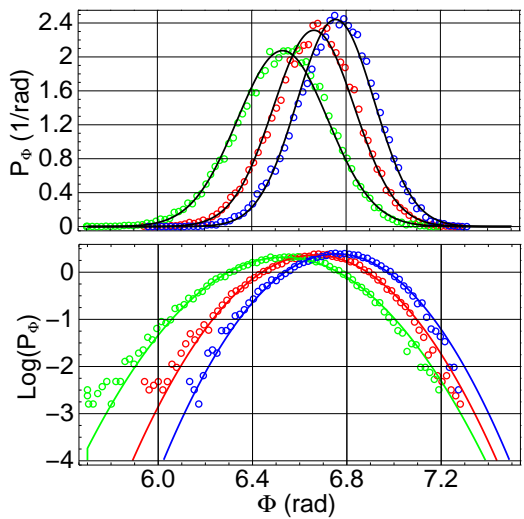}}
\caption{\label{Fpbdhs0} Color online.
The normalized probability density $P_\Phi$ obtained with different
applied torques. Form left to right, the MD results are shown for
$\tau$= -20, 0, and +20 pN$\cdot$nm by green, red, and blue points,
respectively. The solid lines exhibit analytical distributions
\Req{Pphi} corresponding to the measured values of $l_t$ and
$\Phi_\tau$. The lower panel displays the same data in
semi-logarithmic coordinates.
}\end{figure}

\Rfg{pbdhs0} shows the probability distributions $P_\Phi$ for the
GC-alternating sequence for three representative values of $\tau$.
All of the distributions are close to the analytical Gaussians defined
by \Req{Pphi} with different $l_t$.  Since the width of the bells
changes, the neat shapes of the computed distributions are not due to
the harmonicity of the torsional potential. These Gaussian
shapes result from the central limit theorem of the probability theory
whatever the underlying potential is. As seen in \Rfg{pbdhs1}, the
single-step twist fluctuations at GpC and CpG steps produce wide and
skewed non-Gaussian distributions strongly different from that
predicted by \Req{U} (see Appendix).  With the
temperature around 300K, the local DNA dynamics goes far beyond the
area where the harmonic approximation is valid.  However, the
torsional fluctuations of four consecutive bps already give an almost
ideal Gaussian. It can be formally described by \Req{U} and
\req{Pphi}, but the shape of this bell does not correspond to the
harmonic approximation of the local free energy. The Gaussian profile
of fluctuations in long DNA is linked with the single-step
distributions by a linear growth of the variance with the chain
length. Consequently, not just the apex zones of the skewed
distributions in \Rfg{pbdhs1}, but their entire shapes contribute.
Therefore, the anharmonicity is significant, but hidden. In addition,
the twist fluctuations at consecutive steps are anticorrelated and
partially cancel out.

\begin{figure}[ht]
\centerline{\includegraphics[width=7cm]{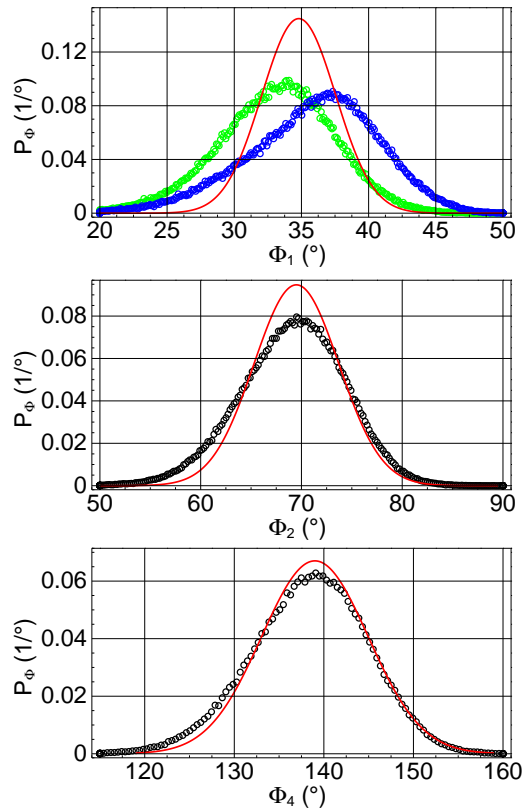}}
\caption{\label{Fpbdhs1} Color online.
The probability density $P_\Phi$ for GC-alternating fragments of one,
two and four bps (from top to bottom) obtained with $\tau=0$. The
solid red lines exhibit the analytical distributions \Req{Pphi}
corresponding to the measured values of $l_t$ and $\Phi_0$. On the
top panel, the distributions for GpC and CpG steps are shown in green
(left) and blue (right), respectively.
}\end{figure}

The asymmetry of the single-step PMFs is the probable cause of the
variable torsional stiffness of the GC-alternating fragment.  In the
first approximation, the $l_t$ value is proportional to the second
derivative of the PMF in the energy minimum (see \Req{U}). For an
asymmetric PMF a decrease in $l_t$ may be expected when the external
torque pushes towards the even slope of the energy profile. In the
GC-alternating sequence both single step distributions are left-skewed
(see \Rfg{pbdhs1}); so the right-hand slope of the PMF is steeper than
the opposite one, which explains the sign of the trend in $l_t$
observed in \Rfg{alvstq}. The nearly flat $l_t$ profile for the
AT-alternating fragment can be also rationalized because in this case
a strong positive skewness of TpA steps is partially compensated by a
negative skewness of ApT steps (see Appendix). Preliminary analysis of
other sequences reveals that the strong negative skewness of the CpG
single-step distributions is exceptional (see Appendix). The homopolymer
ApA and GpG steps are nearly symmetrical whereas the single-step
distributions for AG- and AC-alternating DNA indicate that they would
behave similarly to the AT-alternating fragment. These conclusions
should be yet verified in more intensive computations, but we expect
that for random DNA the macroscopic torsional stiffness should be
nearly constant because among the steps with skewed distributions
positive and negative skewness are equally represented. In contrast,
for short sequence motives anharmonic effects of both signs are
possible. They can be very significant because biological systems
operate with much larger torques than we use here. For instance,  the
binding sites of the phage 434 repressor contain a variable 4 bp
spacer that does not interact with the protein and supposedly
participates in gene control via the sequence-dependent elasticity
\cite{Hogan:87}.  In the complexed state, this spacer is always
overtwisted by about 30$^\circ$ \cite{Koudelka:06}, that is ten times
the amplitude of twisting in \Rfg{alvstq}.

The experimental bending rigidity of free DNA is characterized by
$l_b\approx50$ nm \cite{Hagerman:88}. The measured $l_t$ values vary
between 36 and 109 nm depending upon the specific methods and
conditions \cite{Fujimoto:06}. Observations of sequence effects are
rare \cite{Fujimoto:90}, and there are a few reports on the influence
of supercoiling \cite{Song:90b,Selvin:92,Naimushin:94}. If we assume that MD
overestimates the stiffness of DNA uniformly then the convergent
estimate of $l_t$ is around 90 nm, close to its value in single
molecule experiments \cite{Strick:99,Bryant:03}. The bias can be due
to the neutralizing salt condition in MD or other factors (see
Appendix). The nearly quantitative agreement between MD and
experiment is remarkable because none of the parameters used in
simulations was adjusted to reproduce the DNA elasticity. We hope,
therefore, that the detailed microscopic picture provided by MD
captures the qualitative physical trends dictated by the atom-level
mechanics of the double helix. Our results argue that, under normal
temperature, the local DNA elasticity is strongly anharmonic.
Extrapolation from the apparent harmonic behavior of macroscopic DNA
is not justified despite a good agreement with atomistic simulations
for chain lengths beyond one helical turn \cite{Mzbj:06,Mzprl:07}. In
addition, these computational observations shed new light upon some
earlier controversial issues.

According to \Rfg{alvstq}, with the helical twist slightly shifted
from the equilibrium value the sequence dependence of the DNA
elasticity can be significantly changed and enhanced. The measured
torsional stiffnesses are similar without applied torque, but diverge
with untwisting. The deformability of DNA is long considered as a
possible governing factor in the sequence-specific site recognition
\cite{Hogan:87}, but this mechanism requires strong sequence
dependence of elastic parameters compared to that observed in
experiments with free DNA \cite{Fujimoto:90}. As we see the properties
of the relaxed DNA cannot be simply transfered to supercoiled and/or
protein bound DNA states. Additional studies are necessary to check if
the elastic properties of the specific binding sites change under
torsional stress. Its magnitude may be very large in some protein
complexes \cite{Koudelka:06}.

Another debated issue concerns the mechanisms of gene regulation via
DNA supercoiling \cite{Peter:04,Travers:05a}. Many of such
observations are readily rationalized if we assume that the sensitive
promoters are regulated via the torsional stiffness.  Even a slight
shift in its value has a dramatic effect on the probabilities of
strong twisting fluctuations.  Many transcription factors are designed
to bind the double helix at two sites separated by a spacer of several
base pair steps. They can work as sensors of torsional fluctuations in
DNA. A strong twisting fluctuation may be necessary for binding such
factor or for recognition by other proteins of a permanently bound
torsional sensor.  \Rfg{pbdhs0} shows that for fluctuations observable
during 164 ns, physiological modulations of the torsional stress would
change the corresponding probabilities by several times.  For less
frequent larger fluctuations the effect would be much stronger. One
can extrapolate the pattern in \Rfg{pbdhs0} to events observable in
the millisecond time range, and this leads to essentially
all-or-nothing switching.

The external torque shifts the distributions in \Rfg{pbdhs0} by
changing symmetrically the energies of opposite fluctuations. If the
shape of the distributions does not change each pair of curves should
give a single intercept between the corresponding two apexes.
However, if the shifting is accompanied by widening, one more intercept
should appear in the range of large twisting opposite to the torque
direction. For instance, the negative torque shifts the distribution
in \Rfg{pbdhs0} to the left, but the simultaneous widening raises its
right wing and, with very large overtwisting, the left curve should go
above the other two. It is seen in \Rfg{pbdhs0} that the vertical
difference between between the three plots indeed exhibits a reducing
trend with large $\Phi$. This effect is somewhat paradoxical and it
qualitatively contradicts the behavior of simple models where the
torsional energy depends upon a single variable. Our attempts to
reproduce it in discrete wormlike chains with anharmonic torsion
potentials were unsuccessful. However, such behavior is possible, in
principle, due to coupling between different degrees of freedom and it
requires further studies.

To conclude, it appears that small external torques can significantly
alter the torsional stiffness of the double helical DNA. The effect is
sequence-dependent, and, under variable degree of supercoiling,
different stretches of the double helix can become locally softer or
stiffer. This can represent a versatile mechanism of gene regulation via
the probabilities of strong twisting fluctuations.

\section{Appendix}
\setcounter{figure}{0}
 \captionsetup{labelformat=empty,labelsep= none,
  justification=centerlast,width=.9\columnwidth,aboveskip=5pt}

\subsection*{Simulation protocols}

Tetradecamer DNA fragments were modeled with AT-alternating
(d(AT)$_7$) and GC-alternating (d(GC)$_7$) sequences.  The starting
states for classical MD simulations were prepared as follows. The
solute in the canonical B-DNA conformation \cite{Arnott:72} was
immersed in a 6.2-nm cubic cell with a high water density of 1.04. The
box was neutralized by placing Na$^+$ ions at random water positions
at least 5 \AA\ from the solute. The system was energy minimized and
dynamics were initiated with the Maxwell distribution of generalized
momenta at low temperature. The system was next slowly heated to 293 K
and equilibrated during 1.0 ns. After that the water density was
adjusted to 0.997 by removing the necessary number of water molecules
selected randomly at least 5 \AA\ from DNA and ions, and the
simulations were continued with NVT ensemble conditions. The
temperature was maintained by the Berendsen algorithm
\cite{Berendsen:84} applied separately to DNA, water, and ions,  with
a relaxation time of 10 ps.  Simulations with external torques started
from equilibrated states after a few nanoseconds of free dynamics.

The AMBER98 forcefield parameters \cite{Cornell:95,Cheatham:99} were
used with the rigid TIP3P water model \cite{Jorgensen:83}.  The
electrostatic interactions were treated by the SPME method
\cite{Essmann:95}, with the common values of Ewald parameters, that is
9 \AA\ truncation for the real space sum and $\beta\approx 0.35$.  To
increase the time step, MD simulations were carried out by the
internal coordinate method (ICMD), \cite{Mzjcc:97,Mzjchp:99} with the
internal DNA mobility limited to essential degrees of freedom and
rotation of water molecules and internal DNA groups including only
hydrogen atoms slowed down by weighting of the corresponding inertia
tensors.  \cite{Mzjacs:98,Mzjpc:98} The double-helical DNA was modeled
with all backbone torsions, free bond angles in the sugar rings, and
rigid bases and phosphate groups. The effect of these constraints is
insignificant, as was previously checked through comparisons with
standard Cartesian dynamics \cite{Mzjacs:98,Mzbj:06}. The time step
was 0.01 ps and the DNA structures were saved every 5 ps. All
trajectories were continued to accumulate 2$^{15}$ points, i.e. to
about 164 ns. To verify the results for d(GC)$_7$ three additional
trajectories were computed with torques $\tau$= -20, 0, and +20
pN$\cdot$nm, respectively These computations were carried out in
parallel on 32 processors starting from independent equilibrated
states.  In all simulations the B-DNA conformations were well
conserved without visible slow trends or accumulated deformations.
\Rfg{titc0}S shows some standard time plots for two representative
trajectories. It is seen that the overall properties of the double
helices remain stable and that the DNA structures were
well-equilibrated before the beginning of the production runs.

\begin{figure}[ht]
\centerline{\includegraphics[width=8cm]{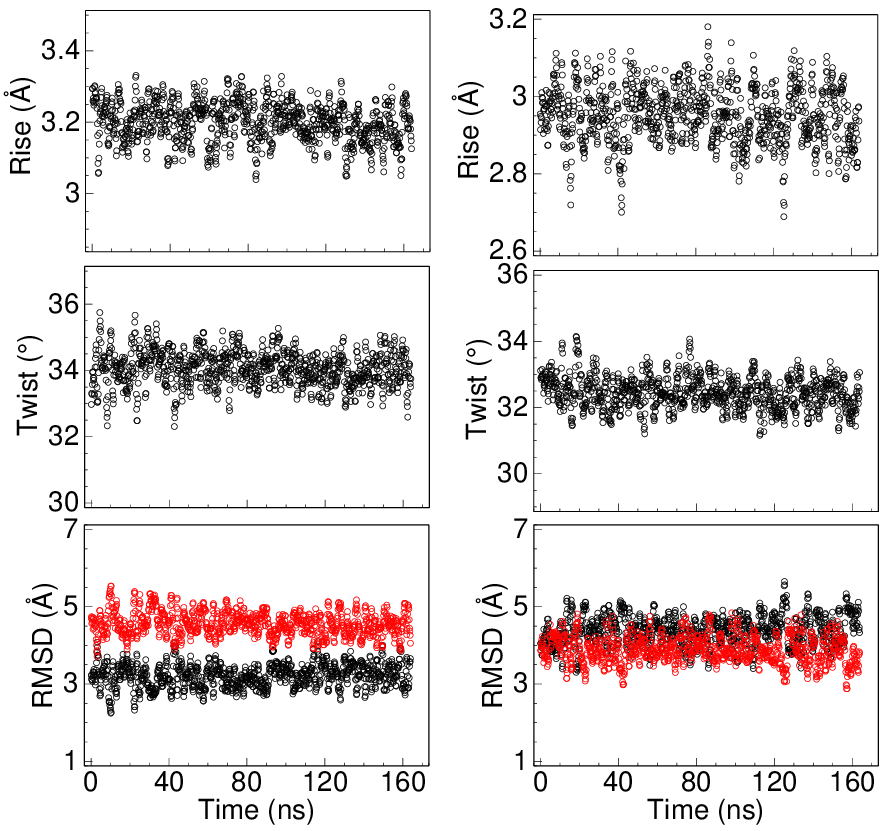}}
\caption{ FIG. \rfg{titc0}S: \footnotesize\label{Ftitc0}
The time evolution of the double helical structures during MD
simulations with applied torque $\tau$= -20 pN$\cdot$nm. The results
for the GC- and AT-alternating sequences are shown in the left and
right columns, respectively. The helical parameters are averages for
the central 12 bp obtained by program 3DNA \cite{Lu:03c}. The all-atom
RMSDs from the canonical A-DNA (red) and B-DNA (black) were computed
for the entire fragment length (14 bp). The analysis was carried out
for 1000 states equally spaced throughout the trajectories and all
plots were smoothed with a sliding window of 82 ps.
}\end{figure}

The choice of the fragment length and sequences is consistent with the
recent computations \cite{Mzjpc:08,Mzjctc:09,Mzjpc:09} and it was
dictated by the following considerations. The length slightly larger
than one helical turn is convenient for measuring the elastic
parameters of DNA \cite{Mzjpc:09}. These molecules are homopolymers of
AT- and GC-units, therefore, they cannot have distinguished asymmetric
structures like static bends. True homopolymer DNA duplexes have
special features and, in free MD with the AMBER forcefield, these
structures deviate from the canonical B-DNA stronger than AT- and
GC-alternating sequences \cite{Mzjctc:05}. The terminal base pairs
open rather frequently during nanosecond time scale MD, which
significantly perturbs the flanking DNA structure. Because this
dynamics cannot be averaged during the accessible duration of MD
trajectories, we blocked it by applying non-perturbing upper distance
restraints as explained elsewhere \cite{Mzjctc:09}. The statistical
convergence of MD sampling also suffers from rare transitions of
backbone torsion angles to non-standard states considered as
forcefield artifacts \cite{Dixit:05a,Perez:07a}. In the present
simulations such transitions occurred mainly in the terminal base pair
steps excluded from analysis. No $\alpha/\gamma$ flips were observed
in the middle d(CG)$_6$ stretches. A few such transitions that occurred
in the d(TA)$_6$ fragments were documented and discussed in our
previous report \cite{Mzjctc:09}.  They increased the statistical
noise in the measured parameters, but did not cause overall structural
perturbations.  \Rfg{dapbd0}S shows several representative
distributions of the backbone torsions involved in non-canonical
transitions \cite{Dixit:05a}. The number of $\alpha/\gamma$ flips in
d(TA)$_6$ was maximum two per trajectory, but some of them were
reversed.  The distribution with the maximal population of
non-canonical states shown in \Rfg{dapbd0}S corresponds to a
trajectory with a single stable flip.

\begin{figure}[ht]
\centerline{\includegraphics[width=8cm]{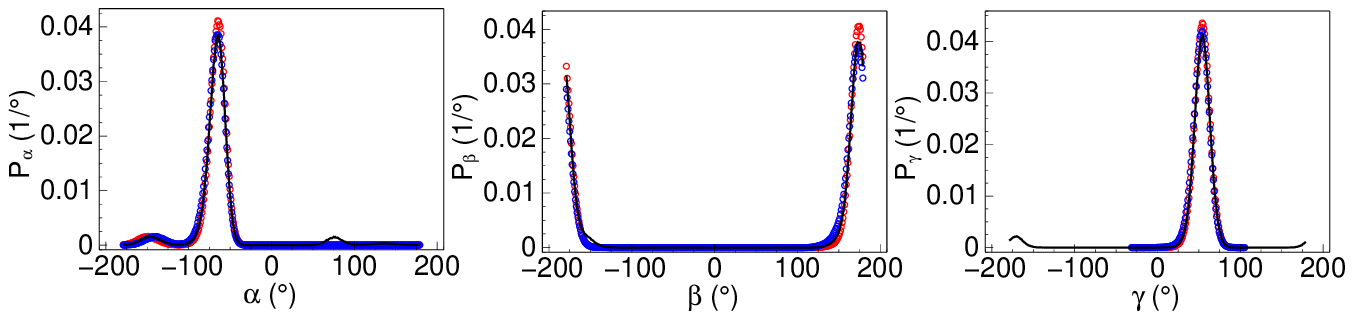}}
\caption{ FIG. \rfg{dapbd0}S: \footnotesize\label{Fdapbd0}
Distributions of $\alpha$, $\beta$, and $\gamma$ backbone torsions for
three representative trajectories. Plots shown by open circles
correspond to the data in \Rfg{titc0}S, with the results for d(TA)$_6$
and d(CG)$_6$ in red and blue, respectively. The solid black line
shows the distributions for a trajectory with the largest population
of non-standard conformers ($\tau$= +4 pN$\cdot$nm, d(TA)$_6$).
}\end{figure}

The sampled conformations of the double helix were analyzed by program
3DNA \cite{Lu:03c}, with only 11 central base pair steps considered
(d(TA)$_6$ and d(CG)$_6$). Accurate direct measurement of the length
of short double helices encounters technical difficulties
\cite{Lu:99,Mzjpc:09}. Therefore, the corresponding DNA length was
computed as 11*0.335 nm, that is assuming the length of one step
corresponding to experiment. The error in the attributed length can be
partially responsible for the systematic bias in the absolute values
of other measured parameters, but it does not affect qualitative
trends.

\subsection*{External torques}

Steady external torsional load was applied as described in detail in
our recent report \cite{Mzjctc:09}. This method distributes forces
over selected groups of atoms and compensates them by reactions
applied to other atoms so as to zero the total external force and
torque. Because the forces are applied at different points internal
stress and deformations are introduced that correspond to overall
twisting. The method was thoroughly verified in Brownian dynamics
simulations of calibrated discrete wormlike chain models
\cite{Mzjctc:09}. Notably, it was checked that the torque values in
the range of interest cause linear elastic responses in perfect
agreement with theoretical predictions and negligible side effects.

\subsection*{Evaluation of statistical errors}

Evaluation of errors in MD simulations is based upon the following
assertions from the probability theory. Consider a random variable $x$
with expectation ${\bf M}x=\xi$ and variance ${\bf D}x=\sigma^2$. We
can take $n$ samples of $x$ and compute
$$
\overline{x}=\frac1n(x_1+x_2+...+x_n)=\frac1n\sum^n_{k=1}x_k
$$
and
$$
S^2=\frac1{n-1}\sum^n_{k=1}(x_k-\overline{x})^2
$$
called the sample average and variance, respectively. Both
$\overline{x}$ and $S^2$ are random variables, with ${\bf
M}\overline{x}=\xi$ and ${\bf M}S^2=\sigma^2$, i.e. $\overline{x}$
and $S^2$ provide unbiased estimates of $\xi$ and $\sigma^2$,
respectively. It is also known that
\beq\label{EDx}
{\bf D}\overline{x}=\frac{\sigma^2}n
\eeq
and, if $x$ is a Gaussian random variable,
\beq\label{EDS}
{\bf D}S^2=2\sigma^4/(n-1).
\eeq
\Req{Dx} and \req{DS} are used for evaluation of statistical
errors.

\begin{figure}[ht]
\centerline{\includegraphics[width=6cm]{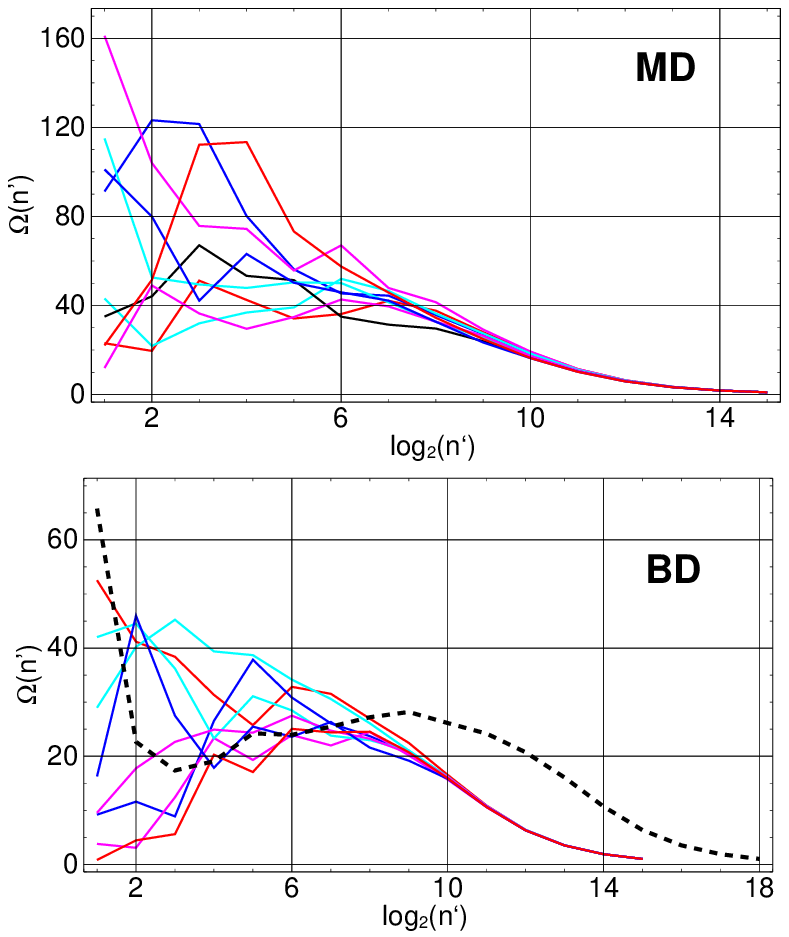}}
\caption{ FIG. \rfg{blker}S: \footnotesize\label{Fblker}
Analysis of statistical errors by the method of block averages
\cite{Frenkel:96}. The upper panel displays the results for the nine
MD trajectories of the GC-alternating fragment. For comparison, the
lower panel presents a similar analysis of eight equivalent BD
trajectories of the same DNA fragment. The equivalence means that MD
and BD trajectories have identical durations and saving intervals. The
BD simulations were carried out under zero torque by using a discrete
WLC model from earlier reports \cite{Mzjpc:09,Mzjctc:09}. The black
dashed line on the lower panel displays the results for a single eight
times longer BD trajectory.
}\end{figure}

In our MD simulations the random variable is the winding angle of one
helical turn, $\Phi$, with expectation ${\bf M}\Phi$ and variance
${\bf D}\Phi$. The torsional persistence length is computed as
$$
l_t=L/{\bf D}\Phi 
$$
where $L$ is the DNA length. It can also be obtained from the shifts in
${\bf M}\Phi$ caused by external torques of different magnitude.  The
true values of ${\bf M}\Phi$ and ${\bf D}\Phi$ are estimated by using,
respectively, the sample average and variance computed over all $n$
points saved during an MD trajectory. However, \Req{Dx} and \req{DS}
cannot be applied straightforwardly because they are valid only for
statistically independent samples, i.e. the time intervals between the MD
states must be suitably large compared to the torsional relaxation time.
The data saving interval is commonly much smaller, therefore, the errors
are evaluated by using the method of block averages \cite{Frenkel:96}.
The trajectory is successively divided in $n'=2,4,...,2^{15}$ stretches
(blocks) and the sample variances $S'^2$ are computed by using $n'$ block
averages instead of individual samples. When the blocks are longer than
the torsional relaxation time the block averages are independent and
$S'^2/n'\approx const=\sigma^2/\tilde n$, where $\tilde n$ is the effective
number of independent samples provided by the trajectory. This value
should be used in place of $n$ in \Req{Dx} and \req{DS}. In practice, it
is convenient to draw the plots of
$$
\Omega(n')=\frac{nS'^2}{n'\sigma^2},
$$
with respect to $\log_2n'$. When statistical independence is reached,
such plots exhibit a plateau with $\Omega\approx n/\tilde n=\tau^c$,
which gives the required estimate of $\tilde n$. Parameter $\tau^c$
is the effective correlation time measured in trajectory saving steps.

Analysis of the MD data for the GC-alternating DNA is shown in
\Rfg{blker}S. For benchmark comparison, we also present a similar
treatment of Brownian dynamics (BD) trajectories of a discrete
wormlike chain (WLC) model used in our recent studies
\cite{Mzjpc:09,Mzjctc:09}.  With $n'$ decreasing, the plots display
emergence of a plateau and the growth of statistical noise. From these
plots the $\tau^c$ values are estimated as 50 and 30 for MD and BD,
respectively, in agreement with the corresponding relative rates of
torsional relaxation \cite{Mzjpc:09}. The $\tau^c\approx50$ gives
$\tilde n$=655 and the relative error of 5.5\% in the measured $l_t$
values. This accuracy is sufficient for our purposes. The lower panel
of \Rfg{blker}S also displays the improvement that might be obtained
with longer trajectories.  As expected, with $2^3$ longer BD
trajectory the plateau is less noisy, but its value is similar.

\subsection*{Sequence-dependent distributions of twist fluctuations}

\begin{figure}[ht]
\centerline{\includegraphics[width=8cm]{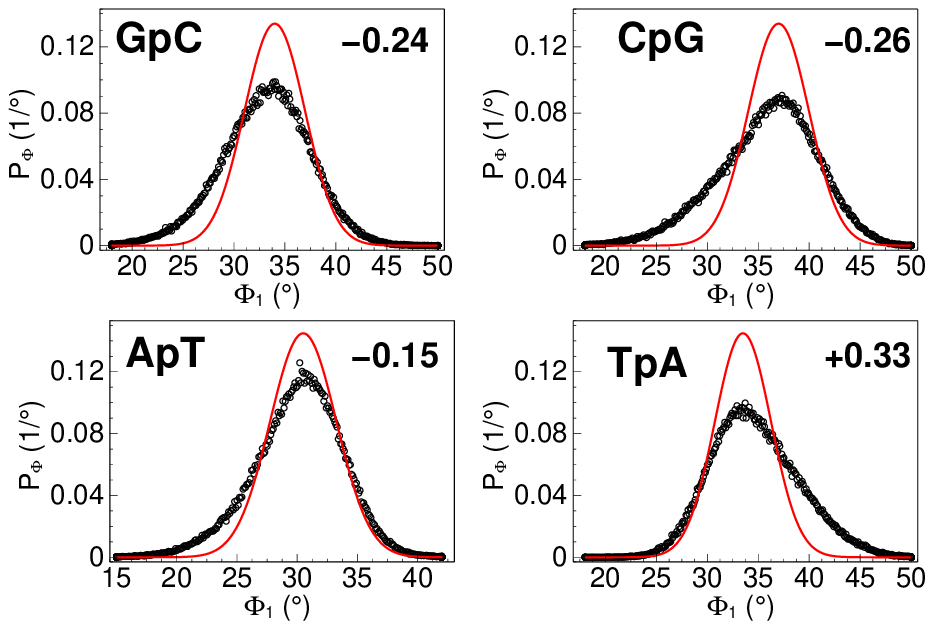}}
\caption{ FIG. \rfg{pbd1sp}S: \footnotesize\label{Fpbd1sp}
Normalized single-step probability densities of torsional
fluctuations in MD. The red lines show the Gaussian distributions
corresponding to the harmonic WLC model with the corresponding values
of $l_t$. The Gaussians were shifted to the maxima of the computed
distributions.  The MD data shown are from the trajectories with zero
applied torques.
}\end{figure}

\Rfg{pbd1sp}S compares the observed single step distributions of
torsional fluctuations in d(AT)$_7$ and d(GC)$_7$ with the
corresponding distributions in equivalent coarse grained models, i.e.
discrete WLC models with elastic parameters identical to those in MD.
The numbers shown in the upper right corners display the
corresponding mode skewness computed as
$$
\rm (mean - mode)/(standard\ deviation).
$$
Similar results for other base pair steps are shown in \Rfg{al1sp}S.
These data were obtained by MD simulations of tetradecamer fragments
with the corresponding alternating or homopolymer sequences. Duration
of all trajectories was about 8 ns.

\begin{figure}[ht]
\centerline{\includegraphics[width=7cm]{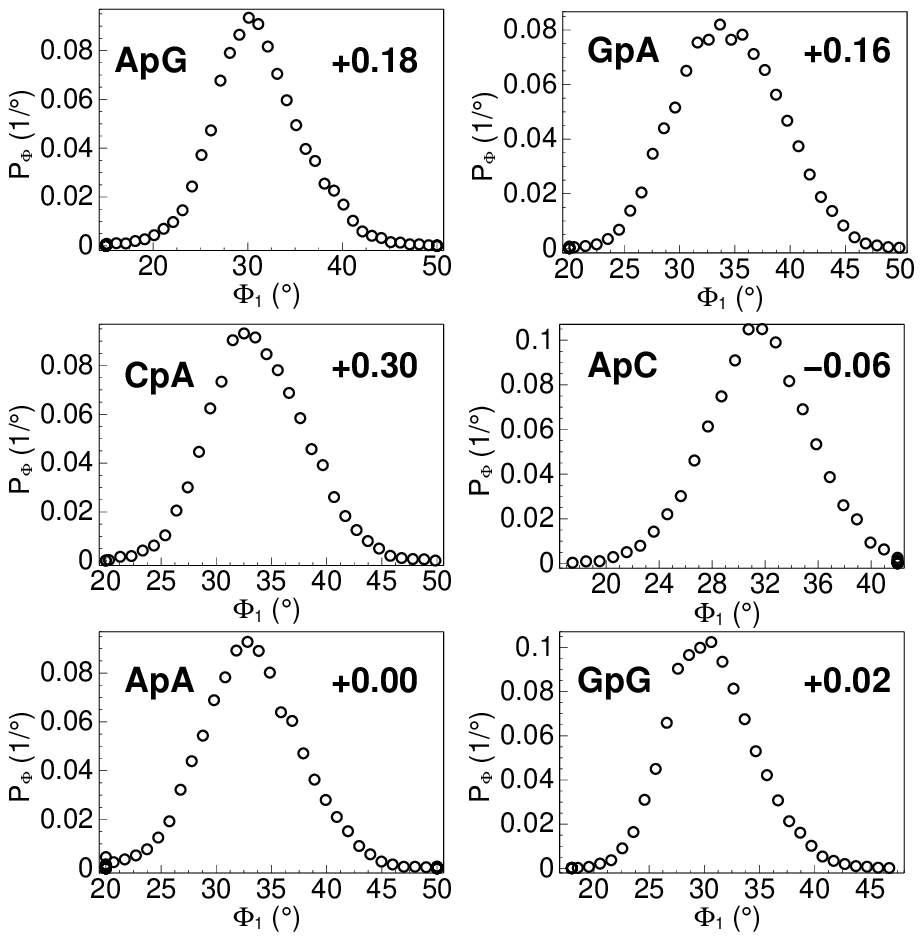}}
\caption{ FIG. \rfg{al1sp}S: \footnotesize\label{Fal1sp}
Normalized single-step probability densities of torsional
fluctuations for other sequences.
}\end{figure}

\bibliography{mzpaper}

\end{document}